\documentclass[prx,amsmath,amssymb,aps,twocolumn,superscriptaddress,showpacs,notitlepage,floatfix,bibnotes]{revtex4-2}
\usepackage{dsfont}
\usepackage{yfonts}
\usepackage{bm}
\usepackage{mathrsfs}
\usepackage{graphicx}
\usepackage{verbatim}
\usepackage{wasysym}
\usepackage{xcolor}
\usepackage{xspace}
\usepackage{tikz}
\usepackage{physics}
\usepackage{floatrow}
\usepackage{diagbox}
\usepackage{footmisc}
\usepackage[normalem]{ulem}
\usepackage[caption=false,label font={bf,normalsize}]{subfig}
\AtBeginDocument{\tabcolsep=7pt} 
\usepackage{booktabs}

\newcommand{\comments}[1]{}

\newlength{\fighskip} \fighskip=2pt
\newlength{\figvskip} \figvskip=3pt

\usepackage{xspace}
\renewcommand{\ket}[1]{\vert{#1}\rangle}
\renewcommand{\bra}[1]{\langle{#1}\vert}

\renewcommand{\braket}[2]{\langle#1|#2\rangle}

\newcommand{\Z}{\mathbb{Z}}

\newcommand{\Ft}{\ensuremath{\widetilde{F}}}
\newcommand{\Zt}{\ensuremath{\widetilde{Z}}}
\newcommand{\Xt}{\ensuremath{\widetilde{X}}}

\makeatletter
\def\l@subsubsection#1#2{}
\makeatother

\usepackage[pdfpagelabels,pdftex,bookmarks,breaklinks]{hyperref}
\definecolor{darkblue}{RGB}{0,0,127} 
\definecolor{darkgreen}{RGB}{0,150,0}
\hypersetup{
	colorlinks, 
	linkcolor=darkblue, 
	citecolor=darkgreen, 
	filecolor=red, 
	urlcolor=blue,
  pdfauthor = {Brenden Roberts, Sagar Vijay, Ashvin Vishwanath, Arpit Dua},
  pdftitle = {Topological invariants of Floquet codes}
}

\begin{document}
\title{Geometric phases in generalized radical Floquet dynamics}

\author{Brenden Roberts}
\thanks{brendenroberts@fas.harvard.edu}
\affiliation{Department of Physics, Harvard University, Cambridge, MA 02138, USA}
\author{Sagar Vijay}
\thanks{sagar@physics.ucsb.edu}
\affiliation{Department of Physics, University of California, Santa Barbara, CA 93106, USA}
\author{Arpit Dua}
\thanks{adua@caltech.edu}
\affiliation{Department of Physics and Institute for Quantum Information and Matter, California Institute of Technology, Pasadena, California 91125, USA}

\date{\today}
  
\begin{abstract}
The Pancharatnam phase is a generalization of the Berry phase that applies to discrete sequences of quantum states. 
Here, we show that the Pancharatnam phase is a natural invariant for a wide class of quantum many-body dynamics involving measurements.
We specifically investigate how a non-trivial Pancharatnam phase arises in the trajectories of Floquet quantum error-correcting codes and show that this phase can be extracted in a ``computationally-assisted"  interferometry protocol, involving additional post-processing based on the measurement record that defines a given quantum many-body trajectory. 
This Pancharatnam phase can also be directly related to the Berry phase accrued by continuous \emph{unitary} evolution within a gapped phase. 
For the $\Z_2$ Floquet code of Hastings and Haah, we show that the associated family of unitary evolutions is the radical chiral Floquet phase.
We demonstrate this correspondence explicitly by studying an exactly-solvable model of interacting spins. 
\end{abstract}
\maketitle

Insights from quantum information theory have recently been instrumental in uncovering universal features in out-of-equilibrium quantum dynamics~\cite{Fisher_2023_review}.
Ongoing developments in quantum simulation have driven interest in dynamics, which broaden the paradigm of unitary evolution, for instance, incorporating measurements, classical processing, and quantum feedback.

A particularly interesting example is quantum cellular automata (QCA), the class of locality-preserving unitary transformations that includes finite-depth local quantum circuits~\cite{Farrelly_2020}.
Intriguingly, one-dimensional shift QCA describing chiral translations can emerge naturally on the boundaries of a system subject to periodic driving referred to as anomalous Floquet unitary evolution~\cite{po2016chiral,harper2017floquet}.
One remarkable example is realized by a finite-depth circuit on a bosonic system, which induces a period-doubled translation of a fermion at the edge. This boundary action is referred to as the fermionic shift QCA, and the unitary is referred to as the \emph{radical chiral Floquet} (RCF) phase~\cite{po2017radical,potter2017dynamically,fidkowski2019interacting}.
It is marked by a bulk automorphism of topological superselection sectors and its unconventional fractional edge mode, demonstrating a bulk-boundary correspondence.
Under deviations from the ideal case, we note that an assumption of many-body localization (MBL) via coupling to strong disorder is necessary to avoid heating to an incoherent high-temperature state~\cite{po2016chiral,po2017radical,potter2017dynamically}.

Recent work implements such boundary translations via sequences of measurements specifying a two-dimensional error-correcting code referred to as the Floquet code.
In particular, for the $\Z_2$ Floquet toric code (FTC) of Ref.~\cite{hastings2021dynamically}, operators in a Majorana algebra associated with the restriction of bulk stabilizers at the boundary exhibit an automorphism and, due to the breaking of reflection symmetry, manifest a chiral propagation of fermion modes on the edge~\cite{aasen2023measurement}.
More generally, a variety of Floquet quantum error-correcting codes are emerging~\cite{davydova2023quantum,Davydova_2023,bauer2023topological,dua2023engineering,kesselring2022anyon,ellison2023floquet,XcubeFloquet_2022,townsendteague2023floquetifying}, along with other kinds of quantum dynamics with projective measurements in which a known logical subspace is preserved~\cite{lavasani2021topological,lavasani2023monitored,sriram2023topology}.
While these have diverse features, each fundamentally describes an ensemble of trajectories composed of states obtained sequentially via projective measurement.
For Floquet quantum codes, the dynamically evolving logical information is retrievable through a known labeling of each state in the sequence based on its logical content. 

In this correspondence, we explore the geometric phases of ``generalized radical Floquet dynamics,'' encompassing both the RCF unitary~\cite{po2017radical} and the
$\Z_2$ FTC~\cite{hastings2021dynamically}.
Notably, both examples perform the same bulk automorphism of topological superselection sectors.
We demonstrate that this automorphism induces a quantized geometric phase of $\pi$, manifesting as a topological contribution to the geometric Berry or Pancharatnam phase of a single period of the unitary or measurement sequence, respectively.

We provide an interferometry protocol for the Pancharatnam phase of the measurement dynamics by utilizing the inherent error-correcting capability of the Floquet code. 
Adapting to the inherent randomness in measurement results necessitates implementing outcome-dependent unitary transformations. 
We contend that this is feasible through quantum error correction without introducing extra phase accumulation, thus constituting an example of a computationally-assisted observable~\cite{lee2022measurement,weinstein2023nonlocality}.

Extending our analysis to systems with boundaries, we find that sequences of geometric phases accrued through the steps of measurement dynamics can serve as indicators for anomalies in measurement schedules. For instance, in the $\Z_2$ FTC with a boundary, certain 3-round (6-round) schedules yield non-cyclic (cyclic) sequences of geometric phases. We posit this cyclic nature as a requirement for constructing measurement schedules for planar variants of Floquet codes.

\textbf{Bulk invariant for $\Z_2$ Floquet toric code.---}
A sequence of (pairwise non-orthogonal) wavefunctions $\mathcal C = \{ \ket{\psi_0},\ket{\psi_1}, \ket{\psi_2},\ldots,\ket{\psi_M}\}$ admits a \emph{Pancharatnam phase}
\begin{equation}
\eta_{\mathcal C} = \Im\log\braket{\psi_0}{\psi_1}\braket{\psi_1}{\psi_2}\ldots\braket{\psi_M}{\psi_0}~.
\end{equation}
Below we demonstrate that if $\mathcal C$ is a Floquet code trajectory, $\eta_{\mathcal C}$ contains a universal contribution, which can be robustly extracted.

Adiabatic time evolution under the Schr{\"o}dinger equation produces (up to dynamical phase) parallel transport satisfying $\Im\bra{\psi(t)} \frac{d}{dt} \ket{\psi(t)}=0$, and the holonomy of this connection is the Berry phase~\cite{berry1984quantal,simon1983holonomy}.
Although holonomy is not directly applicable to a discrete sequence, the Pancharatnam phase of a pair of non-orthogonal states is known to be equal to the integral of the connection along any member of a family of continuous ``null-phase curves'' (NPC)~\cite{mukunda2003bargmann}, which includes Fubini-Study geodesics~\cite{samuel1988general}.
By repeated concatenation of such curves, the Berry phase of a continuous piecewise-NPC (CPNPC) connecting the states in a discrete sequence is seen to be equivalent to the Pancharatnam phase of the sequence itself.
We will see that CPNPC curves are implemented by unitary Floquet evolutions.

We consider the $\Z_2$ FTC on the honeycomb lattice with periodic boundary conditions and edges tri-colored as red, green, or blue. 
The code is defined using a repeated sequence of two-qubit measurements: one first measures the $X\otimes X$ checks on all red edges, then $Y\otimes Y$ checks on green edges, and finally $Z\otimes Z$ checks on blue edges.
After a warm-up stage, in each subsequent round, one obtains the instantaneous stabilizer group (ISG) of a toric code concatenated with 2-qubit repetition codes on measured edges.
The logical information evolves deterministically and is hence preserved~\cite{hastings2021dynamically}.

Under the $\Z_2$ FTC, logical string operators undergo nontrivial orbits given by $\{\Zt_a,\Zt_b\} \leftrightarrow \{\Xt_b,\Xt_a\}$, where $a$ and $b$ label the logical qubits~\cite{hastings2021dynamically}.
The unitary action on the codespace is thus
given by
\begin{equation}
\mathcal M[\phi] = e^{i\phi}(H_a\otimes H_b)\circ\mathrm{SWAP}~,
\label{eq:code_action}
\end{equation}
with $H_a$, $H_b$ Hadamards acting on the logical qubits.
This dynamical effect reflects the homotopy of gapped linear interpolations of stabilizer fixed-point Hamiltonians~\cite{aasen2022adiabatic} and has been shown to be robust to errors~\cite{vu2023measurement}.

The action $\mathcal M[\phi]$ is one realization of the bulk automorphism exchanging $e$ and $m$ superselection sectors; others---e.g.~a Hadamard on each logical qubit---would likewise be consistent with the automorphism.
We propose that a full dynamical signature is given by the phase eigenvalues and eigenvectors of the unitary.
The eigenvectors of $\mathcal M[\phi]$ are eigenstates of the ``inner logical'' operators $\Ft_{c_1} = \Zt_a \Xt_b$ and $\Ft_{c_2} = \Xt_a \Zt_b$ which transport a fermion around the cycles of the torus, denoted here by $c_1$ and $c_2$~\footnote{These are referred to as inner logical operators because they are non-local stabilizers from the group of check operators that become logical operators in the ISGs.}.
The eigenvalues are a triply degenerate $e^{i\phi}$, as well as a unique $e^{i(\phi+\pi)}$ with eigenstate $\ket{\Ft_{c_1} = -1 , \Ft_{c_2} = -1}$.
While the specific unitary leading to the bulk $e$-$m$ automorphism may vary, such a dynamical signature can be shown to be nontrivial for all such unitaries.

The observation of the phase eigenvalue structure of $\mathcal M[\phi]$ is impeded by the generic angle $\phi$ (in the absence of errors $\phi=0$ for the $\Z_2$ FTC~\cite{Geometric_Floquet_SM}).
To overcome this, we propose an interferometry experiment to extract the universal contribution.
Unfortunately, this introduces another obstacle: due to a combination of errors and ``gauge inequivalence'', identical states undergoing the same measurement-based evolution are very likely to be orthogonal afterward.
Here, errors can be understood as Pauli noise creating anyons, and by gauge inequivalence we refer to the random measurement outcomes of the checks leading to a difference in the underlying 2-qubit repetition code stabilizer eigenvalues upon completing a full cycle.
We avoid postselection, however, by appealing to error correction for the former. For the latter, we observe that using the knowledge of the measurement outcomes, a shallow Pauli circuit can be written to match the gauge of eigenvalues of the 2-qubit checks without imparting an additional phase.

The interferometry protocol for the universal phase contribution is as follows:
\begin{enumerate}
\item Prepare code states $\ket u = \ket{\Ft_{c_1}=-1,\Ft_{c_2}=-1}$ and $\ket v = \mathcal O \ket u$, where $\mathcal O$ is a symmetry operator.
\item Evolve each state under a single round of measurements: $\{\ket u,\ket v\}\mapsto \{\ket{u'},\ket{v'}\}$.
\item Perform error correction and gauge (eigenvalue) matching via classical processing: 
$\{\ket{u'},\ket{v'}\}\mapsto \{\ket{u''},\ket{v''}\} \equiv \{W_{u}U_{u}\ket{u'}, W_{v}U_{v}\ket{v'}\}$.
\item Apply $\mathcal O$ and compute the overlap
\begin{equation}
\Im\log\bra{u''}\mathcal O\ket{v''} = \pm1~,
\end{equation}
with ($-1$) $+1$ signifying (non-)trivial evolution.
\end{enumerate}
For an error rate below the maximum likelihood threshold, the homology class of single-qubit errors can be obtained by inferring the syndromes from the measurement outcomes~\cite{hastings2021dynamically}.
Errors are thus removed via syndrome-dependent unitary transformations $U_{u}$ and $U_{v}$ applied to the two states.
Additional unitaries $W_{u}$ and $W_{v}$, determined by classical processing, are applied to the states to enforce the agreement of stabilizer eigenvalues after a full cycle.
(Such Pauli circuits generically need not contribute to the overall phase.)
The final states $\ket{u''}$ and $\ket{v''}$ share both a stabilizer group and logical state. 
We identify the nontrivial relative phase $e^{i\pi}=-1$ accumulated on this code state as the $\Z_2$ topological invariant of the class of automorphism dynamics.

\begin{figure}[ht]
\includegraphics[width=\columnwidth]{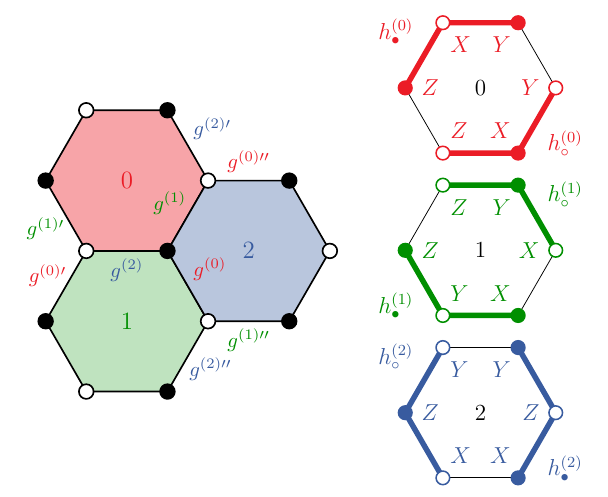}
\caption{\label{fig:radical_unitary_conventions} On the left is shown the unit cell of Eq.~\eqref{eq:floquet_H}.
The plaquette stabilizer integrals of motion are $W^{(0)}_p = \prod_{j \in \partial p} X_j$, $W^{(1)}_p = \prod_{j \in \partial p} Y_j$, and $W^{(2)}_p = \prod_{j \in \partial p} Z_j$, with $p$ labeling an appropriate plaquette.
The commuting local Hamiltonian terms appearing in the exponential in $U_F$ are shown on the right.
Each can be recognized under the exact solution as a short fermionic string between sites on a sublattice.
}
\end{figure}

\textbf{Relationship to noninteracting invariants.---}
The universal contribution to the Pancharatnam phase of the Floquet code both signifies its automorphism dynamics and computes the related CPNPC holonomy.
We present an explicit unitary describing such a CPNPC in the RCF phase of Ref.~\cite{po2017radical}; i.e., it implements a nontrivial fermionic QCA in one dimension~\cite{potter2017dynamically,fidkowski2019interacting}.
This continuous unitary should be compared to the quantum circuits of Refs.~\cite{aasen2022adiabatic,sullivan2023floquet}.
While the construction of a bulk invariant for such a phase remains challenging, the trajectories associated with the $\Z_2$ FTC permit calculation of the universal geometric phase from the adiabatic evolution of a band insulator.

We define the following Floquet unitary with period $T$ on the honeycomb lattice: $U_F \equiv U(t=T)$, where $U(t) = \mathcal T \exp[-\frac{3\pi i}{4T}\int_0^t H_F(t')\,dt']$ and
\begin{equation}
H_F(t) = \sum_{r=0,1,2} \kappa_r(t) \sum_{\bm s} \left( h^{(r)}_{\circ,\bm s} - h^{(r)}_{\bullet,\bm s} \right),
\label{eq:floquet_H}
\end{equation}
where $\bm s$ runs over six-site unit cells as shown in Fig.~\ref{fig:radical_unitary_conventions}, along with the local terms.
The time dependence is given by~\footnote{While localizing on-site disorder is conventionally introduced in an additional step, we are concerned with eigenstates of an integrable model; our results are stable under translation symmetry breaking if MBL is introduced.}
\begin{equation}
(\kappa_0(t),\kappa_1(t),\kappa_2(t)) = \begin{cases}(0,0,1)~,&0 \leq t < \frac T3\\(1,0,0)~,&\frac T3 \leq t < \frac{2T}{3}\\(0,1,0)~,&\frac{2T}{3} \leq t \leq T\end{cases}~.
\label{eq:floquet_H_2}
\end{equation}

The Floquet operator $U_F$  with period $T$ is the product of constant $t=\frac T3$ evolutions $U_r$ under stabilizer Hamiltonians $H_{F,r} = \sum_{\bm s} ( h^{(r)}_{\circ,\bm s} - h^{(r)}_{\bullet,\bm s})$, $r=0,1,2$.
Plaquette operators factorize the many-body Hilbert space into static flux configurations $W^{(r)}_{\bm s} = \pm1$~\cite{kitaev2006anyons,po2017radical,hastings2021dynamically}.
Each of the $H_{F,r}$ can be exactly solved using emergent Majorana fermion operators~\cite{kitaev2006anyons}, and within a flux sector, gauge fixing leads to a spinless BdG Hamiltonian in symmetry class D.
The classification of constant unitary evolution is the same as that of static Hamiltonians, so one finds a $\Z$ invariant, with each $U_r$ being a member of the trivial class~\cite{roy2017periodic}.

The $\Z_2$ FTC satisfies the null-phase condition because each term in Eq.~\ref{eq:floquet_H} fluctuates maximally within the trajectories of the $\Z_2$ FTC: that is, $\bra{\psi_j}h^{(r)}_{\circ,\bm s}\ket{\psi_j} = \bra{\psi_j}h^{(r)}_{\bullet,\bm s}\ket{\psi_j} = 0$ for all $j$.
Consequently $\bra{\psi(t)}\frac{d}{dt}\ket{\psi(t)} = -\frac{3\pi i}{4 T} \bra{\psi_j}U_r^\dag(t)H_{F,r} U_r(t) \ket{\psi_j} = -\frac{3\pi i}{4 T} \bra{\psi_j}H_{F,r}\ket{\psi_j} = 0$.
Moreover, we note that code states are equal-weight superpositions of all $U_r$ Floquet eigenstates, so the Berry phase $\Phi$ of the full evolution computes the topological index of the unitary on the circle:
\begin{equation}
\Phi = i \int dt\,\braket{\psi(t)}{\dot\psi(t)} = i \int dt \Tr U_F^{-1} \partial_t U_F~.
\end{equation}

It is convenient to describe adiabatic paths under the Floquet unitary via another Hamiltonian
\begin{widetext}
\begin{align}
H[J_0,J_1,J_2] &= -\sum_{\bm s} \Big[\sqrt{J_0} g^{(0)}_{\bm s} + \tau(J_0,J_2,J_1) g^{(0)\prime}_{\bm s} + \tau(J_0,J_1,J_2) g^{(0)\prime\prime}_{\bm s} + \sqrt{J_1 J_2} (K^{(0)}_{\bm s} + L^{(0)}_{\bm s}) \nonumber\\
&\hspace{1cm} + \sqrt{J_1} g^{(1)}_{\bm s} + \tau(J_1,J_2,J_0) g^{(1)\prime}_{\bm s} + \tau(J_1,J_0,J_2) g^{(1)\prime\prime}_{\bm s} + \sqrt{J_2 J_0} (K^{(1)}_{\bm s} + L^{(1)}_{\bm s}) \nonumber\\
&\hspace{1cm} + \sqrt{J_2} g^{(2)}_{\bm s} + \tau(J_2,J_1,J_0) g^{(2)\prime}_{\bm s} + \tau(J_2,J_0,J_1) g^{(2)\prime\prime}_{\bm s} + \sqrt{J_0 J_1} (K^{(2)}_{\bm s} + L^{(2)}_{\bm s}) + J^W \sum_r W_{\bm s}^{(r)}\Big],
\label{eq:tc_H}
\end{align}
\end{widetext}
with $\tau(x,y,z) = x^{\left(\frac{2y+z}{2y+2z}\right)}$ and operator conventions as shown in Fig.~\ref{fig:radical_unitary_conventions}.
The $K^{(r)}$ and $L^{(r)}$ are, for example,
\begin{equation}
\begin{gathered}
K^{(0)} = i g^{(1)\prime} h^{(0)}_\bullet = i W^{(0)} g^{(2)\prime} h^{(0)}_\circ~,\\
L^{(0)} = -i g^{(1)\prime} h^{(0)}_\circ = -i W^{(0)} g^{(2)\prime} h^{(0)}_\bullet~,
\end{gathered}
\end{equation}
with the others defined equivalently.
We take the limit $J^W\to\infty$, projecting into the trivial global flux sector $W_{\bm s}^{(r)}=+1$. 
The point $H^{(0)}$ at $(J_0,J_1,J_2)=(1,0,0)$, is the commuting projector model of a toric code on the triangular superlattice of 0-plaquettes, and similarly for $H^{(1)}$ and $H^{(2)}$~\cite{hastings2021dynamically}.

Eq.~\eqref{eq:tc_H} satisfies, as $t$ runs from 0 to $\frac T3$,
\begin{equation}
H(t) \equiv U(t) H^{(0)} U^\dag(t) = H[J_0=1-\tfrac{3t}{T},J_1=\tfrac{3t}{T},J_2=0]~.
\end{equation}
That is, the simplex $J_0+J_1+J_2=1$ reproduces, on the $J_2=0$ boundary, the adiabatic limit of the Floquet time evolution of $H^{(0)}$ to $H^{(1)}$.
One sees that $H^{(0)}$ is mapped to itself under $U_F$ but circumnavigates the boundary, always remaining a gapped commuting projector model.
The existence of such a Hamiltonian is due to the nature of the Floquet unitary $U_F$ which satisfies an MBL condition $(U_F)^n=1$ for some constant $n$. This condition and Lieb--Robinson bounds constrain the growth of the time-evolved local stabilizers; see~\cite{Geometric_Floquet_SM} for details.

Description of the interior of the simplex, $J_0+J_1+J_2=1$, relies on the exact solution using Majoranas~\cite{kitaev2006anyons}.
Within a flux sector, gauge fixing chooses the logical content of a spinless BdG Hamiltonian with time reversal and charge conjugation symmetries.
The code state is the Fock space vacuum of all positive-energy modes, or equivalently the filled state of all negative-energy modes; the many-body geometric phase derives from the cumulative single-particle phases as
\begin{equation}
\Phi_\mathrm{vac} = \frac12 \sum_{\bm k,a} \Phi^-_{\bm k,a} = \frac12 \sum_{\bm k,a} \bra{\psi_{\bm k,a}} \partial_t\ket{\psi_{\bm k,a}}~,
\end{equation}
where $\Phi^-_{\bm k,a}$ are the Berry phases of filled single-particle bands labeled by momentum $\bm k$ and band index $a$~\cite{Geometric_Floquet_SM}. 
Though the single-particle gap closes at $\bm k=0$, this is accidental with respect to the symmetry, and there is no phase associated with winding in momentum space; rather, a contour's $\Z_2$ classification is determined by its winding around the gapless point in parameter space.
Even if time-reversal symmetry is broken, the gapless point is simply displaced from $\bm k = 0$, and the path retains its $\Z_2$ topological invariant, associated with the adiabatic pumping of fermion parity~\cite{teo2010topological}.
This index manifests as a universal contribution to the many-body geometric phase~\cite{Geometric_Floquet_SM}.

As a consequence of free fermion integrability, $U_F$ describes an adiabatic path not just for the ground state but for every eigenstate.
All eigenstates accrue a topological phase contribution identical to the measurement trajectories in the $\Z_2$ FTC; the difference is that many measurement trajectories are possible given an initial state, whereas the unitary describes a single path for each eigenstate.
These are nevertheless identified by constructing Pauli circuits for error correction and matching gauge measurement outcomes in the Floquet code, as discussed previously.

\begin{figure}[ht]
\includegraphics[width=\columnwidth]{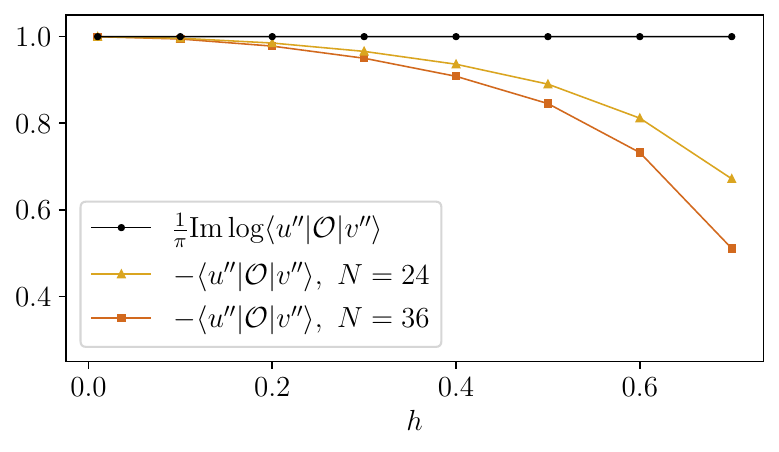}
\caption{\label{fig:perturbed_unitary}The outcome of a unitary interferometry protocol is shown to produce a universal contribution to the geometric phase even away from the stabilizer fixed point, as shown for toruses on $N$ qubits.
A field $-h \sum_i Z_i$ is applied on the superlattice, and variational ground states $\{\ket u,\ket v\}$ on finite systems are evolved under $U_F$.
Though the overlap diminishes due to the perimeter law scaling of Wilson loops, all points having a finite modulus exhibit the phase invariant signature of the automorphism.
}
\end{figure}

In this section, we have shown that trajectories of the $\Z_2$ FTC are in correspondence with adiabatic curves in the time-reversal invariant free-fermion model Eq.~\eqref{eq:tc_H} and that the quantized phase invariant is robust to deformations preserving integrability, including those corresponding to Pauli noise.
While the measurement code (even with errors) produces stabilizer states, we show in Fig.~\ref{fig:perturbed_unitary} that the phase invariant remains observable under more generic perturbations.
Away from the stabilizer fixed point, the superlattice Wilson loop $\mathcal O = \prod_{i \in c} Z_i$ for a nontrivial cycle $c$ scales with a perimeter law, but always maintains the nontrivial dynamical signature in the sign~\footnote{The magnetic field in Fig.~\ref{fig:perturbed_unitary} can be considered a type of correlated error, and shows that the invariant is observable beyond the strictly correctable regime, if expectation values of homological loops remain finite.}.
That is, we do not compute the dressed logical operator but simply measure the bare operator even away from the stabilizer fixed point.

\textbf{Boundary anomaly and geometric phases.---}
The construction of a planar variant of the $\mathbb{Z}_2$ FTC necessitates a 6-round schedule, contrasting the 3-round schedule which is applicable to a torus topology~\cite{haah2021boundaries,Paetznick_2023,Gidney_2022}.
Due to this, when constructing gapped boundaries for the Floquet code, we posit a crucial prerequisite: achieving a null geometric phase after a single measurement cycle within the Floquet code. This, in turn, mandates the return of the ISG to its exact initial configuration after a certain number of cycles. In the context of the planar $\mathbb{Z}_2$ Floquet code, the requisite recurrence does not occur with the 012 schedule; however, it is appropriately realized using a rewinding schedule, specifically the 012102 pattern~\cite{dua2023engineering}.

In the planar Floquet code with the 012 schedule, a charge creation operator is added as a measurement check to create a particular gapped boundary condensate.
However, under the measurement sequence, this condensation operator evolves into a non-local stabilizer that is the inner logical operator, thus inducing a transition from an arbitrary logical state to an eigenstate of the logical operator~\cite{hastings2021dynamically}.
Subsequent measurements preserve this state as the checks commute with the logical operator.
Conversely, the rewinding schedule keeps the logical information intact, facilitating a return to the same ISG and logical state after a single period, albeit subject to environmental errors.

For the rewinding schedule, the sequence of geometric phases, derived through our interferometry protocol that includes error correction, also traces back to a null geometric phase following a single period of the rewinding schedule.
Consequently, the anomaly associated with boundary construction can be effectively explored by examining the progression of geometric phases accrued during the measurement dynamics.

\textbf{Discussion.---}
We have identified a quantized geometric phase imparted to code states in Floquet codes, which serves as a robust observable signature of a bulk automorphism by focusing on the $\Z_2$ FTC as an illustrative example.

This phase signature establishes a compelling equivalence between the dynamic measurement schedules and the continuous unitary trajectories within the Hilbert space. 
In the precise scenario considered, the toric code duality automorphism identifies the Floquet code as a generalized manifestation of what is known as radical chiral Floquet-enriched topological order.
That is, we extend the unitary classification to the case also involving measurements.
Whereas the chiral edge index of the Floquet phase can be computed directly from the unitary, the bulk invariant of the code is predicated on encoded qubits in the state, as well as an error rate below the threshold.
Thus, the dynamical classification of Floquet code orders forms a subset of the unitary classification but carries the advantage of evident bulk invariants observable in code state trajectories.

The equivalence can be extended to scenarios involving boundaries.
However, in the presence of boundaries, fermionic chiral edge modes lead to stabilizers at the edge following an orbit of complexity proportional to $O(\partial V)$.
This implies that the MBL condition $(U_F)^n=1$ is not fulfilled. 
Consequently, it is not clear how to formulate a representative model akin to Eq.~\eqref{eq:tc_H}; however, using the rewinding schedule this is again possible.
If one excludes boundary terms from the Hamiltonian, the system remains generically gapless, as evidenced by a substantial degeneracy associated with the edge degrees of freedom.
This lack of a gap precludes the construction of adiabatic unitary paths in this context.

This reasoning is consonant with the established boundaries for the $\Z_2$ FTC.
Notably, these boundaries are characterized by a modified six-round bulk schedule which serves to nullify the automorphism.
Conversely, Floquet codes not implementing a bulk automorphism---such as the CSS Floquet code---accommodate boundaries without alterations to the bulk schedule, sharply distinguishing them from their bulk-automorphism counterparts.

The geometric Pancharatnam phases discussed in this work provide a way to benchmark the fidelity of the measurement processes used to implement Floquet codes in practice.
They may also be useful for more general studies of circuit dynamics, such as those involving both unitaries and measurements.
For instance, this framework must apply to possible occurrences of automorphisms in other dynamically generated codes, such as random stabilizer codes in one dimension~\cite{darmawan2022low}.
Quantum many-body dynamics with randomly-located measurements can also give rise to dynamically-protected codespaces in which the logical operators are well-understood~\cite{lavasani2021topological,lavasani2023monitored,sriram2023topology}.
The Pancharatnam phase for these dynamics remains to be investigated.

\textbf{Acknowledgement.---} 
The authors wish to acknowledge discussions with Tyler Ellison, Ruihua Fan, Nat Tantivasadakarn, Ruben Veressen, DinhDuy Vu, and Carolyn Zhang.
We especially thank Ashvin Vishwanath for significant suggestions and involvement.
B.R. is supported by the Harvard Quantum Initiative Postdoctoral Fellowship in Science and Engineering. S.V. acknowledges the support of the Alfred P. Sloan Foundation through a Sloan Research Fellowship. 
A.D. is supported by the Simons Foundation through the collaboration on Ultra-Quantum Matter (651438, AD) and the Institute for Quantum Information and Matter, an NSF Physics Frontiers Center (PHY-1733907).
The authors acknowledge that part of this work was performed at the Aspen Center for Physics, (which is supported by National Science Foundation grant PHY-1607611) and at the KITP program on Quantum Many-Body Dynamics and Noisy Intermediate-Scale Quantum Systems (which is supported in part by the National Science Foundation grants PHY-1748958 and PHY-2309135).

\bibliography{refs}

\end{document}


\title{Supplemental material: ``Geometric phases in generalized radical Floquet dynamics''}
\date{\today}
\maketitle

\section{Geometric phase of Floquet codes}
\subsection{Pancharatnam phase of commuting measurements}

Consider a pure state on $N$ qubits stabilized by $\S = \ev{S_1,S_2,\dots,S_N} \subset \mathcal P$.
Following a projective measurement of a Pauli operator $\tilde Q$, we obtain the $+1$ eigenstate of stabilizer $\tilde Q$ or $-\tilde Q$ depending on whether the measurement outcome is $+1$ or $-1$.
We treat $\tilde Q$ or $-\tilde Q$ as our resulting stabilizer $Q$ with eigenvalue $+1$.
That is, while it may be that either $Q \in \S$ or $q \not\in \S$ before measurement, we cannot have $-Q \in \S$.
A single measurement taking stabilizer group $\S$ to $\S'$ is associated with the Pancharatnam phase $\eta_Q = \Im\log\braket{\S}{\S'}\braket{\S'}{\S}=0$.

One round of a Floquet code entails the sequential measurement of mutually commuting Pauli operators $\bm Q=\{Q_1,Q_2,\ldots,Q_M\}$, with Pancharatnam phase
\begin{equation}
\eta_{\bm Q} = \Im\log \braket{\S_0}{\S_1}\braket{\S_1}{\S_2} \cdots \braket{\S_M}{\S_0} = \sum_{\alpha=0}^M \Im\log \braket{\S_\alpha}{\S_{\alpha+1}}~,
\end{equation}
taking $\alpha+1\mod M$.
Considering measurements individually, let a single $Q \not\in \S$ commute with all $S_i$, $i \neq 1$, and anticommute with $S_1$; then the stabilizer group after measurement is $\S' = \ev{S,S_2,\dots,S_N}$.
The contribution to the phase is $\Im\log\braket{\S}{\S'} = \Im\log\bra\S\Pi_Q\ket\S = \Im\log\bra\S(\Pi_Q)^2\ket\S = \Im\log\braket{\S'}{\S'}=0$, where $\Pi_Q$ is the projector onto the $+1$ eigenspace of $Q$.
The remaining term is $\Im\log\braket{\S_M}{\S_0} = \Im\log\bra\S\prod_{j=1}^M\Pi_{Q_j}\ket\S = \Im\log\bra\S\prod_{j=1}^M(\Pi_{Q_j})^2\ket\S = \Im\log\braket{\S_M}{\S_M}=0$,
which relies on the commutation of the $\{Q_j\}$.

Thus, the measurement of commuting Pauli operators never accrues a Pancharatnam phase, so it is equivalent to both a single null-phase curve (NPC) as well as a continuous piecewise-NPC (CPNPC).

\subsection{Pancharatnam phase of \texorpdfstring{$\mathbb{Z}_2$}{Z2} Floquet toric code}
The ISG after round $r$ is $\S^{(r)} = \ev{S^{(r)}_1,S^{(r)}_2,\ldots,S^{(r)}_M}$ with independent generators $S^{(r)}_i$ taken from the gauge group; it is convenient to consider basis states for the code space $\ket{\Ft_{c_1} = \sigma_{c_1}, \Ft_{c_2} = \sigma_{c_2}}$, $\sigma_{c_1,c_2} = \pm 1$, which have rank-1 density matrices $\ket{\text{TC}_r^{(\sigma_{c_1},\sigma_{c_2})}}\bra{\text{TC}_r^{(\sigma_{c_1},\sigma_{c_2})}} = \Pi_r^{(\sigma_{c_1},\sigma_{c_2})}\rho_\infty \Pi_r^{(\sigma_{c_1},\sigma_{c_2})}$, where
\begin{align}
\Pi_r^{(\sigma_{c_1},\sigma_{c_2})} \sim \left(1 + \sigma_{c_1} \Ft_{c_1} \right)\left(1 + \sigma_{c_2} \Ft_{c_2} \right)\prod_i \left(1 + S^{(r)}_i\right).
\end{align}
By $\Ft_{c_1,c_2}$ we denote the inner logical operators of the code~\cite{HH_dynamic_2021}, products of gauge operators along cycles $c_1$ and $c_2$ of the torus, which are invariant under the measurement protocol \footnote{There is an additional sign when written this way: for example, $\Ft_{c_1} = \Zt_a \Xt_b = -\prod_{g^{(r)}\in c_1} g^{(r)}$.}.
The full measurement dynamics is diagonal in this basis, and we do not have to consider the degenerate manifold of code states altogether but can treat the Abelian phase of each basis state separately.

The Pancharatnam phase of the HFC is
\begin{align}
\eta(\sigma_{c_1},\sigma_{c_2}) &=\Im\log\braket{\text{TC}_0^{(\sigma_{c_1},\sigma_{c_2})}}{\text{TC}_1^{(\sigma_{c_1},\sigma_{c_2})}}\braket{\text{TC}_1^{(\sigma_{c_1},\sigma_{c_2})}}{\text{TC}_2^{(\sigma_{c_1},\sigma_{c_2})}}\braket{\text{TC}_2^{(\sigma_{c_1},\sigma_{c_2})}}{\text{TC}_0^{(\sigma_{c_1},\sigma_{c_2})}} \nonumber\\
&= \Im\log\Tr\ket{\text{TC}_1^{(\sigma_{c_1},\sigma_{c_2})}}\braket{\text{TC}_1^{(\sigma_{c_1},\sigma_{c_2})}}{\text{TC}_2^{(\sigma_{c_1},\sigma_{c_2})}}\braket{\text{TC}_2^{(\sigma_{c_1},\sigma_{c_2})}}{\text{TC}_0^{(\sigma_{c_1},\sigma_{c_2})}}\bra{\text{TC}_0^{(\sigma_{c_1},\sigma_{c_2})}} \\
&=\Im\log\Tr\left[\left(1 + \sigma_{c_1} \Ft_{c_1}\right)\left(1 + \sigma_{c_2} \Ft_{c_2}\right) {\prod_p}' \left( 1+W_p\right) {\prod_{g^{(1)}}}' \left( 1+g^{(1)}\right) {\prod_{g^{(2)}}}' \left( 1+g^{(2)}\right) {\prod_{g^{(0)}}}' \left(1+g^{(0)}\right) \right].
\end{align}
The overlap takes the form of a sum of products that contribute only if they act as the identity.
Primed products indicate that a single term of each type---which is not independent due to relationships between the generators---is omitted.
This precludes certain products, namely $\prod_p W_p = 1$ and $\prod g^{(r)} W_p^{(r)} = 1$.
However, other terms remain, which we classify as being either \emph{nonlocal} or \emph{local} depending on whether they include a logical operator or not, respectively.

\subsubsection{Local products}

Local products may include either a finite or an extensive number of terms as long as the weight of each term individually is bounded.
These are generated by $W_p^{(m)} g^{(l)} g^{(l)\prime} g^{(l)\prime\prime} g^{(r)} g^{(r)\prime} g^{(r)\prime\prime} = +1$, where the check operators form the boundary of plaquette $p$ of type $m \neq l,r$.
As an operator equation, the sign applies independent of measurement outcomes.
Some collection of these products can be chosen to generate the group $\mathcal L$ of local terms. 
Elements of $\mathcal L$ always come with an overall $+1$ sign; thus, the phase $\eta$ derives only from the sign structure of nonlocal products.

\subsubsection{Nonlocal products}

Nonlocal products necessarily include at least one of $\{\Ft_{c_1},\Ft_{c_2}\}$.
Elements of the local group, $\mathcal L$, constitute moves we may apply to some nonlocal generators to obtain the full set. Since each move comes with a $+1$, all nonlocal terms related by local moves carry the same sign.
To find the sign of a nonlocal term, note that we can always fix $\Ft_{c_1,c_2}$ in a standard position, defined as a product of checks in a sequence, $\cdots g^{(1)} g^{(2)} g^{(0)} g^{(1)\prime} g^{(2)\prime} g^{(0)\prime} \cdots$ or $\cdots g^{(1)} g^{(0)} g^{(2)} g^{(1)\prime} g^{(0)\prime} g^{(2)\prime} \cdots$.
(These are the minimal units compatible with the coloring, with length three plaquettes.)
Any check operator $g^{(r)}$ appearing in a string operator, anticommutes with two checks also appearing in the operator. 
Consider the first standard form: to bring the product of checks $\prod_{g^{(1)} \in c_1} g^{(1)} \prod_{g^{(2)} \in c_1} g^{(2)} \prod_{g^{(0)} \in c_1} g^{(0)}$ to the form $\Ft_{c_1}$ requires commuting operators through one another.
Note that all anticommuting checks of type $g^{(1)}$ and $g^{(2)}$ are already correctly positioned, so they do not produce a sign.
Each $g^{(0)}$ in the ``interior'' of the product (referring only to the present ordering of terms) anticommutes with a single $g^{(1)}$ operator, and the final $g^{(0)}$ does not move.
For example, for a torus of length 3,
\begin{equation}
(g^{(1)} g^{(1)\prime})(g^{(2)} g^{(2)\prime})(g^{(0)} g^{(0)\prime}) = (g^{(1)} g^{(2)} g^{(1)\prime} g^{(2)\prime})(g^{(0)} g^{(0)\prime}) = -g^{(1)} g^{(2)} g^{(0)} g^{(1)\prime} g^{(2)\prime} g^{(0)\prime}~.
\end{equation}
As the number of ``interior'' $g^{(0)}$ terms is always odd, $\Ft_{c_1}\times\prod_{g^{(1)} \in c_1} g^{(1)} \prod_{g^{(2)} \in c_1} g^{(2)} \prod_{g^{(0)} \in c_1} g^{(0)} = (\Ft_{c_1})^2 = 1$.
This result applies to either standard form, since the products $\prod_{g^{(r)}} g^{(r)}$ commute.
So nonlocal terms involving $\Ft_{c_1}$ or $\Ft_{c_2}$ come with a sign $\sigma_{c_1}$ or $\sigma_{c_2}$, respectively.

The final class of nonlocal products contains $\Ft_{c_1} \Ft_{c_2}$.
We deform the microscopic standard form for the string operators described in the above paragraph in such a way that they intersect only on a single link, and next to this link have a ``local standard form'' $-\Ft_{c_1} = \cdots g^{(0)} g^{(1)} g^{(2)\prime} \cdots$ and $-\Ft_{c_2} = \cdots g^{(2)} g^{(1)} g^{(0)\prime} \cdots$.
The terms can be arranged in the product as
\begin{equation}
\Ft_{c_1} \Ft_{c_2} = (g^{(1)} g^{(2)\prime} \cdots g^{(0)})(g^{(1)} g^{(0)\prime} \cdots g^{(2)}) = (g^{(2)\prime} \cdots g^{(0)})(g^{(0)\prime} \cdots g^{(2)})~,
\end{equation}
noting that $g^{(1)}$ commutes with $\Ft_{c_1}$.
Now the commutation process starting from the ordering
\begin{equation}
\prod_{g^{(1)} \in c_1\ominus c_2} g^{(1)} \prod_{g^{(2)} \in c_1 \ominus c_2} g^{(2)} \prod_{g^{(0)} \in c_1\ominus c_2} g^{(0)}~,
\end{equation}
with $\ominus$ denoting the symmetric difference of the sets of links associated with $c_1$ and $c_2$, proceeds independently for the interior operators of each product, which commute as they are away from the string intersection.
(There may also be $W_p$ terms involved in the deformation from global standard form, but these can be moved freely.)
Commuting the exterior operators involves one additional anticommutation of $g^{(0)}$ and $g^{(2)}$ operators at the end of the product, for an overall sign $-\sigma_{c_1} \sigma_{c_2}$.

Consequently, the Pancharatnam phase of these basis states is
\begin{equation}
\eta(\sigma_{c_1},\sigma_{c_2}) = \Im\log\big(|\mathcal L|(1 + \sigma_{c_1} + \sigma_{c_2} - \sigma_{c_1} \sigma_{c_2})\big).
\end{equation}
That is, $\eta(-1,-1)=\pi$, and $\eta=0$ for the other basis states.

\section{Adiabatic paths for unitary dynamics}

Using a parameterized Hamiltonian, we write adiabatic paths for the unitary RCF dynamics developed in the main text.
Defining, e.g., $U_2(\theta_2) \equiv U(t = \frac{4T}{3\pi}\theta_2)$, $0 \leq t < \frac T3$, to evolve between ISG$_0$ and ISG$_1$, one finds 
\begin{equation}
\begin{gathered}
U_2(\theta_2) g^{(0)} U_2(\theta_2)^\dag = \cos(2\theta_2) g^{(0)} + \sin(2\theta_2) g^{(1)}~,\\
U_2(\theta_2) g^{(0)\prime} U_2(\theta_2)^\dag = \cos(2\theta_2) g^{(0)\prime} + \sin(2\theta_2) g^{(1)\prime}~,\\
U_2(\theta_2) g^{(0)\prime\prime} U_2(\theta_2)^\dag = \cos^2(2\theta_2) g^{(0)\prime\prime} + \sin^2(2\theta_2) W^{(0)} g^{(1)\prime\prime} + i\cos(2\theta_2)\sin(2\theta_2) \left( g^{(0)\prime\prime} h^{(2)}_\bullet - g^{(0)\prime\prime} h^{(2)}_\circ \right)~.\\
\end{gathered}
\end{equation}

The full orbits of the ISG$_0$ stabilizers are $U_F: \{g^{(0)},g^{(0)\prime},g^{(0)\prime\prime}\} \mapsto \{g^{(0)},W^{(0)} g^{(0)\prime\prime},W^{(1)} W^{(2)} g^{(0)\prime}\}$.
The central stabilizer $g^{(0)}$ has period $T$, while the peripheral (primed) stabilizers are exchanged after one Floquet period, acquiring a phase reflecting $\Z_2$ fluxes through the unit cell.
The same is true for ISG$_1$ and ISG$_2$.
The micromotion of the stabilizers is closely related to the radical chiral edge index~\cite{po2016chiral,po2017radical,fidkowski2019interacting}, and to the MBL condition $(U_F)^n=1$ for constant $n$, which as noted in the main text permits writing a commuting projector Hamiltonian.
In a generic flux sector, $n$ equals 4, which corresponds to a period-4 spectral flow of the eigenstates. 
To simplify the construction, we appeal to error correction to remove vortices, which are superlattice anyons of the toric code, and restrict to the vortex-free sector where all stabilizers have a period of at most $2T$.

Based on the above, we seek to write a Hamiltonian $H[\bm J]$ whose parameter space incorporates a path realizing the time evolution of the stabilizers.
Thus, along this path, the model will necessarily be commuting projectors.
Using $\bm J=(J_0,J_1,J_2)$, it is consistent with the simplex condition $J_0+J_1+J_2=1$ to identify $J_0 = \sin^2(2\theta_1) = \cos^2(2\theta_2)$, and similarly for cyclic permutations of $\{0,1,2\}$.
Then the coefficients of the interior checks $g^{(r)}$ are determined to be $\sqrt{J_r}$.
The coefficient of, for example, the peripheral check $g^{(0)\prime}$ on the simplex boundary $J_2=0$ needs to be $\sqrt{J_0}$, while on the boundary $J_1=0$ it must be $J_0$.
For this reason, the function $\tau(x,y,z) = x^{\left(\frac{2y+z}{2y+2z}\right)}$ is used to smoothly interpolate between these on the simplex.
One is ultimately led to the Hamiltonian presented in the main text.

\subsection{Phase diagram}

The exact solution of the spin Hamiltonian utilizes the expanded Hilbert space $X_j = i b^x_j c_j$, $Y_j = i b^y_j c_j$, $Z_j = i b^z_j c_j$, written in terms of Majorana fermion operators $(b^x)^2 = (b^y)^2 = (b^z)^2 = c^2 = 1$.
The physical Hilbert space satisfies the gauge condition $b^x_j b^y_j b^z_j c_j = 1$ for all $j$~\cite{kitaev2006anyons}.
$\Z_2$ variables are assigned to the links of the lattice through fermion parity measurements $u_{ij} = i b^\mu_i b^\mu_j$ on edges $(i,j)$ of type $\mu=x,y,z$.
The logical sector is selected by the gauge fixing of $u_{ij}$, as the logical symmetries manifest as gauge loops around nontrivial cycles of the torus.
The unit cell at position $\bm r$ includes six vertices, which after gauge fixing correspond to three complex fermion modes $d^\mu_{\bm r}$.

\begin{figure}[ht]
\includegraphics[width=0.45\columnwidth]{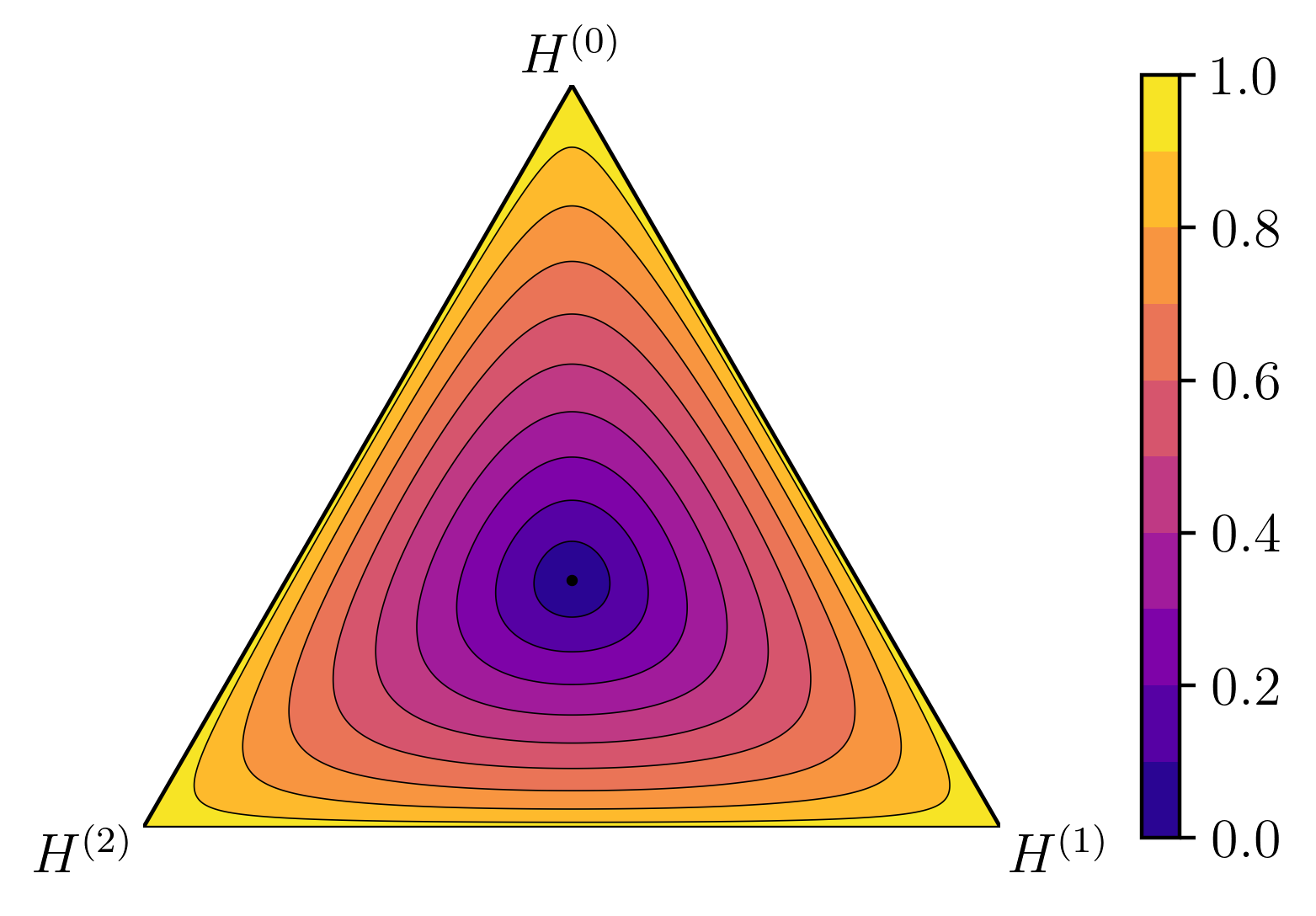}
\hspace{0.5cm}
\includegraphics[width=0.5\columnwidth]{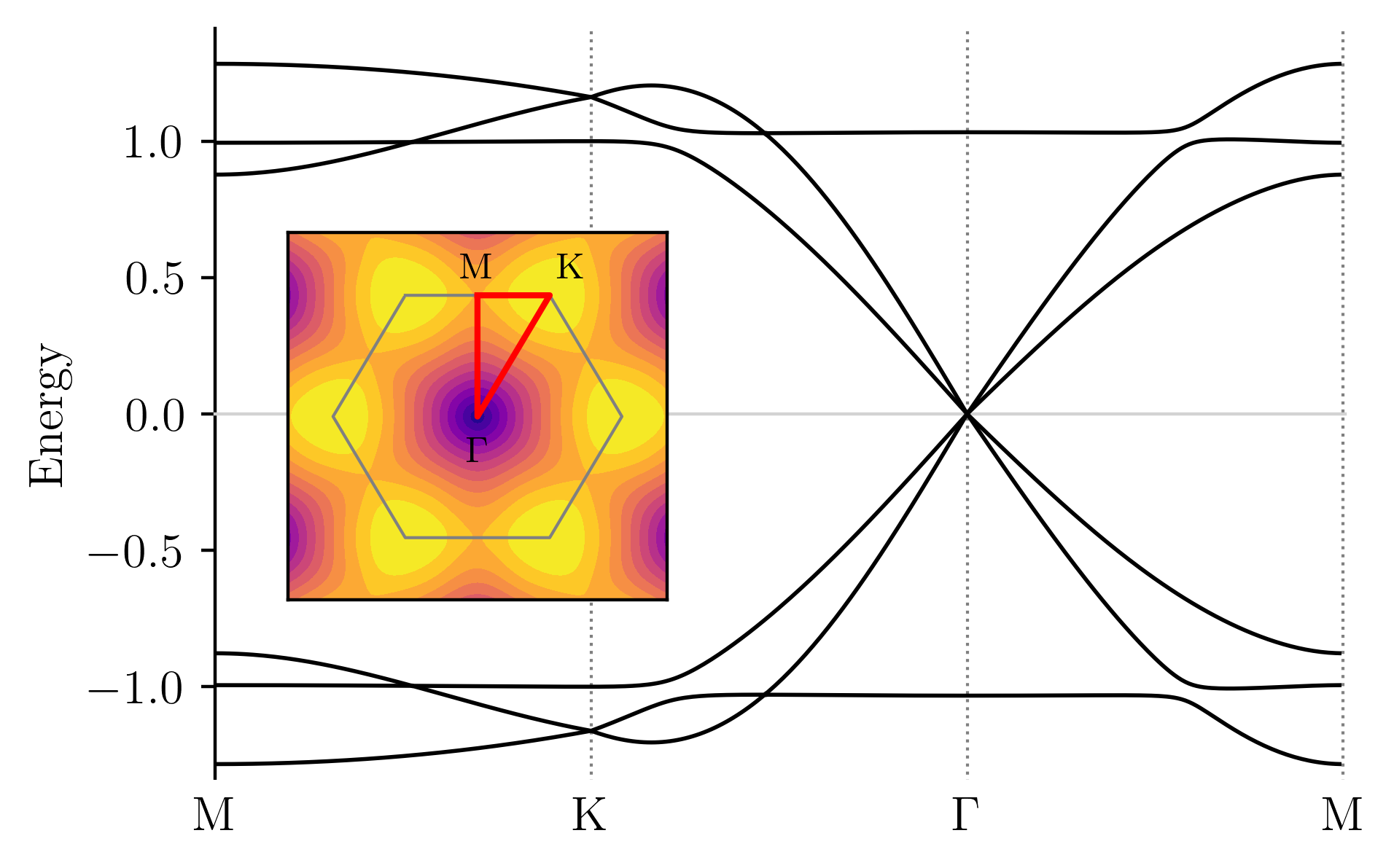}
\caption{\label{fig:bdg_simplex} In the lefthand panel the phase diagram of Eq.~\eqref{eq:H_bdg} is shown, with colors indicating the magnitude of the single-particle gap.
The corners (clockwise from top) are the points $H^{(0)}$, $H^{(1)}$, and $H^{(2)}$, and the path traced by the Floquet unitary time evolution exactly follows the simplex boundary.
The gap closes at a point in the interior, where a double Dirac cone is observed.
The location of the gapless point and the band crossing momentum $\bm k=0$ are not protected by symmetry so can be deformed; however, the gap closing itself cannot be removed.
In the righthand panel, the band structure of the Hamiltonian at the gapless point $(J_0,J_1,J_2)=(\frac13,\frac13,\frac13)$ is shown.
The path traced in the Brillouin zone is shown in the inset.}
\end{figure}

In terms of momentum states $\bm d_{\bm k} = [d^x_{\bm k},d^y_{\bm k}, d^z_{\bm k}]^\top$, the single-particle Hamiltonian is written as
\begin{equation}
H = \sum_{\bm k} [ (\bm d^\dag_{\bm k})^\top,(\bm d_{-\bm k})^\top] \begin{bmatrix} \epsilon_{\bm k} & \Delta_{\bm k} \\ \Delta^\dag_{\bm k} & -\epsilon_{\bm k}\end{bmatrix}\begin{bmatrix}\bm d_{\bm k}\\ \bm d^\dag_{-\bm k}\end{bmatrix} \equiv \sum_{\bm k} \bm D^\dag_{\bm k} M_{\bm k} \bm D_{\bm k}~.
\label{eq:H_bdg}
\end{equation}
The hopping and pairing matrices satisfy $\epsilon_{\bm k}^\dag = \epsilon_{\bm k}$ and $\Delta_{\bm k}^\dag = -\Delta_{\bm k}$, and Eq.~\eqref{eq:H_bdg} has time-reversal and particle-hole symmetries $M^\ast_{\bm k} = M_{-\bm k}$ and $U_C M^\ast_{\bm k} U_C^\dag = -M^\top_{-\bm k}$, respectively.
While the particle-hole symmetry is a consequence of the mapping from spins to fermions, the physical time-reversal symmetry can be broken, taking the model from class BDI to D.
In each case a $\Z_2$ invariant is associated with adiabatic loops, implementing a topological fermion parity pump~\cite{teo2010topological}.
The spectrum is symmetric: $E_{\bm k} = -E_{-\bm k}$, and we denote the unitary diagonalizing $M_{\bm k}$ as
\begin{equation}
U_{\bm k} = \begin{bmatrix}V_{\bm k} & W_{\bm k} \\ W_{\bm k} & V_{\bm k} \end{bmatrix} = \begin{bmatrix}\bm u^+_{\bm k,1} & \bm u^+_{\bm k,2} & \bm u^+_{\bm k,3} & \bm u^-_{\bm k,1} & \bm u^-_{\bm k,2} & \bm u^-_{\bm k,3} \end{bmatrix}.
\end{equation}
The Bogoliubov modes are $\bm \psi_{\bm k} = V^\dag_{\bm k} \bm d_{\bm k} + W^\dag_{\bm k} \bm d^\dag_{-\bm k} = [\psi_{1,\bm k},\psi_{2,\bm k}, \psi_{3,\bm k}]^\top$ with energies $E_{1,\bm k},E_{2,\bm k},E_{3,\bm k} \geq 0$:
\begin{equation}
H = \sum_{\bm k} \bm D^\dag_{\bm k} M_{\bm k} \bm D_{\bm k} =  \sum_{\bm k} \bm D^\dag_{\bm k} \overline V_{\bm k} \Lambda_{\bm k} \overline V^\dag_{\bm k} \overline{\bm d}_{\bm k} 
= \sum_{\bm k} \sum_{a=1,2,3} E_{a,\bm k} \left( \psi^\dag_{\bm k,a} \psi_{\bm k,a} -\frac12\right).
\end{equation}
On the simplex boundary $E_{1,\bm k}=E_{2,\bm k}=E_{3,\bm k}=1$ independent of $\bm k$, as the time evolution $U(t)$ preserves the toric code operator algebra.
The gap closes in the interior; 
the phase diagram on the simplex $J_0+J_1+J_2=1$ is shown in Fig.~\ref{fig:bdg_simplex}.

\subsection{Berry phase}

We write the many-body ground state $\ket\Psi$ as the vacuum of all positive-energy modes $\bm \psi_{\bm k}$~\cite{stone2006fusion}.
Defining $\Gamma_{\bm k} = -(U^\dag_{\bm k})^{-1} V^\dag_{\bm k}$, the following state satisfies $\bm \psi_{\bm k}\ket\Psi=0$ for all $\bm k$:
\begin{equation}
\ket{\Psi} = \frac{1}{\mathcal N} \prod_{\bm k} \exp\left[\frac12 (\bm d^\dag_{\bm k})^\top \Gamma_{\bm k} \bm d^\dag_{-\bm k}\right] \ket 0 = \left(\prod_{\bm k=-\bm k} \frac{1 + \frac12(\bm d^\dag_{\bm k})^\top \Gamma_{\bm k} \bm d^\dag_{\bm k}}{\sqrt{1 + \Tr \Gamma_{\bm k}^\dag \Gamma_{\bm k}}}\right) \left(\prod_{\substack{\bm k>0;\\\bm k\neq-\bm k}} \frac{1 + (\bm d^\dag_{\bm k})^\top \Gamma_{\bm k} \bm d^\dag_{-\bm k}}{\sqrt{1 + \Tr \Gamma_{\bm k}^\dag \Gamma_{\bm k}}}\right) \ket 0~,
\end{equation}
where $\ket0$ is the $d^\mu_{\bm k}$ vacuum and $\bm k > 0$ denotes a set of all momentum states which are not related to each other by inversion.
The Berry connection of the ground state is
\begin{equation}
A = i\bra\Psi d\ket\Psi = \frac i2 \sum_{\bm k} \frac{\Gamma_{\bm k} d \Gamma^\dag_{\bm k} - \Gamma^\dag_{\bm k} d \Gamma_{\bm k}}{1+ \Tr \Gamma^\dag_{\bm k} \Gamma_{\bm k}} = \frac i2 \sum_{\bm k,a} (\bm u^-_{\bm k,a})^\top d \bm u^{-\dag}_{\bm k,a} + d\left[ i \Tr \log U_{\bm k}\right].
\label{eq:many_body_berry}
\end{equation}
Disregarding the total derivative, the Berry connection is the sum of these single-particle states, which are occupied in the ground state.
The factor of $\frac12$ accounts for time reversal pairs $\bm k \leftrightarrow -\bm k$.

The single-particle gap closes at $(J_0,J_1,J_2)=(\frac13,\frac13,\frac13)$ and momentum $\bm k = 0$, where four linearly dispersing bands touch, as shown in Fig.~\ref{fig:bdg_simplex}.
Despite coinciding with an enhanced rotational lattice symmetry, this double Dirac cone is accidental from the perspective of the static Hamiltonian: the symmetry group cannot protect a fourfold degeneracy, and a suitable deformation (which may also entail breaking the time-reversal symmetry) displaces the gap closing.
Accordingly, each band exhibits a trivial Berry phase along contours encircling the gapless point in momentum space; however, along any contour on the simplex $J_0+J_1+J_2=1$ which encircles the gapless point, the $k=0$ states accrue a Berry phase of $\pi$.
The adiabatic paths of the bands of this model are classified by a $\Z_2$ topological invariant that, along with Eq.~\eqref{eq:many_body_berry}, determines the geometric phase of the many-body ground state to be $\pi$ along the adiabatic contour described by $U_F$.
Moreover, because the Berry phase of the positive-energy modes related by particle-hole symmetry is also $\pi$, every Fock state accrues the same topological $\pi$ phase.

\bibliography{refs}